\documentclass[manuscript]{acmart}

\acmConference[CHI '26]{Proceedings of the 2026 CHI Conference on Human Factors in Computing Systems}{April 13--17, 2026}{Barcelona, Spain}
\acmBooktitle{Proceedings of the 2026 CHI Conference on Human Factors in Computing Systems (CHI '26), April 13--17, 2026}
\acmPrice{}

\begin{document}


\title{Not Another EHR: Reimagining Physician Information Needs with Generative AI Technology}

%

\author{Ruican Zhong}
\affiliation{%
  \institution{University of Washington}
  \country{USA}}
\email{rzhong98@gmail.com}

\author{Jicahen Li}
\affiliation{%
  \institution{Northeastern University}
  \country{USA}}

\author{Gary Hsieh}
\affiliation{%
  \institution{University of Washington}
  \country{USA}}

\author{David W. McDonald}
\affiliation{%
  \institution{University of Washington}
  \country{USA}}

\author{Selin S. Everett}
\affiliation{%
  \institution{Stanford University School of Medicine}
  \country{USA}}

\author{Alyssa Unell}
\affiliation{%
  \institution{Stanford University}
  \country{USA}}

\author{Jonathan Carlson}
\affiliation{%
  \institution{Microsoft Research}
  \country{USA}}

\author{Katie Claveau}
\affiliation{%
  \institution{Microsoft Research}
  \country{USA}}

\author{Noel Codella}
\affiliation{%
  \institution{Microsoft}
  \country{USA}}

\author{Khalil Malik}
\affiliation{%
  \institution{Microsoft Research}
  \country{USA}}

\author{Scott Mackie}
\affiliation{%
  \institution{Microsoft}
  \country{USA}}

\author{Eduardo Olvera}
\affiliation{%
  \institution{Microsoft}
  \country{USA}}

\author{Scott Saponas}
\affiliation{%
  \institution{Microsoft Research}
  \country{USA}}

\author{Eric Horvitz}
\affiliation{%
  \institution{Microsoft}
  \country{USA}}

\author{David Rhew}
\affiliation{%
  \institution{Microsoft}
  \country{USA}}

\author{Jim Weinstein}
\affiliation{%
  \institution{Microsoft Research}
  \country{USA}}

\author{Jacob Gross}
\affiliation{%
  \institution{University of Washington School of Medicine}
  \country{USA}}

\author{Amanda K. Hall}
\affiliation{%
  \institution{Microsoft Research}
  \country{USA}}
\email{Amanda.Hall@microsoft.com}

\renewcommand{\shortauthors}{Zhong, et al.}

\maketitle

\section{Introduction}
When we transitioned from paper to EHRs (Electronic Health Records), a patient's chart was no longer compiled in a single paper file at each medical facility where the patient was receiving care~\cite{delrose2000clinical}. While there are many advantages to EHRs (e.g., easily accessible data through digital queries, central storage of medical information, etc.), the sheer volume and complexity of the data presented introduced challenges. Fragmented data interfaces force clinicians to navigate multiple tabs, notes, and systems to piece together a coherent patient picture, which places a significant and growing cognitive load on physicians throughout their daily practice~\cite{walsh2004clinician, gal2021quantifying, chen2014using,mosaly2018relating, murphy2023evolution, bakker2007need, forsythe1992expanding, gonzalez2007information}. Therefore, there is a growing need to rethink how physicians could access and aggregate more easily the information they need.

With the recent rise of generative AI, it is possible to redesign physicians' interactions with data. Specifically, LLMs (Large Language Models) can now summarize and synthesize medical data~\cite{dou2025exploring, wang2025generative}. It can analyze and perform diagnostics~\cite{ahsan2022machine, ghaffar2023evaluation}, make predictions~\cite{bayati2014data, wiens2016patient}, and generate treatment plans~\cite{yu2022role, zhang2023machine}. More importantly, LLMs can now perform UI (user interface) generation~\cite{wu2025improving, gui2025uicopilot, beason2025athena, duan2025automated}, dynamically generating adaptive data visualizations through natural language requests. Thus, it is now feasible to design a generative interface that offers dynamic interactions between the physician and the patient data, where the data presentation is updated according to physicians' requests~\cite{ghosh2024enhancing,
dogiparthi2025building,
neszlenyi2024assistantgpt, zhan2024personalized}. This would enable real-time, personalized interactions, potentially bypassing the current tedious process of collecting and synthesizing the necessary patient information. 

Our position paper also highlights important open questions regarding physicians’ perceptions of AI. Although prior surveys suggest generally positive attitudes toward incorporating AI into clinical workflows,~\cite {fritsch2022attitudes, akinrinmade2023artificial, chew2022perceptions}, it remains unclear how, and under what conditions, physicians would actually trust and rely on AI outputs in practice. And beyond questions of trust, the nature of AI’s engagement with clinicians warrants careful examination: What level of initiative or guidance should AI provide? How should the system present itself, such as posing strictly as a computational tool or as a more anthropomorphized agent? These design choices have significant implications for safety, interpretability, and effective collaboration~\cite{kowalyshyn2025llms, cetinkaya2025between, gerlich2023perceptions}. We argue that understanding and shaping these dynamic interaction patterns should be a central focus in the development of physician facing AI systems. 

This position paper discusses our ongoing research in designing a customizable experience that enables clinicians to more easily retrieve, manage, and synthesize patient data with the support of generative AI. We are interested in: 1) How clinicians navigate patient data; and how their key information needs and challenges shape this process, 2) How clinicians envision generative AI transforming their interaction with patient data with dynamic capabilities, and 3) What design considerations are needed to create clinician-centered generative UI interfaces that support their information needs and workflows. We would like to discuss how AI should be applied in this context to address the current challenges with digital health record systems through dynamic interactions, and explore the potential to design safe and effective generative AI user interface experiences between physicians and data.

\section{Method}
We conducted semi-structured interview sessions with 9 internal physician-technologists at a large tech company. We focused on understanding physicians' information needs for patient care, their pain points, and how they envision generative AI could address these needs. Drawing from these insights, we intend to design an interactive, generative AI user interface system to help physicians work more dynamically with patient data. 

\section{Results}
\begin{figure}[h!]
    \centering
    \includegraphics[width=400pt]{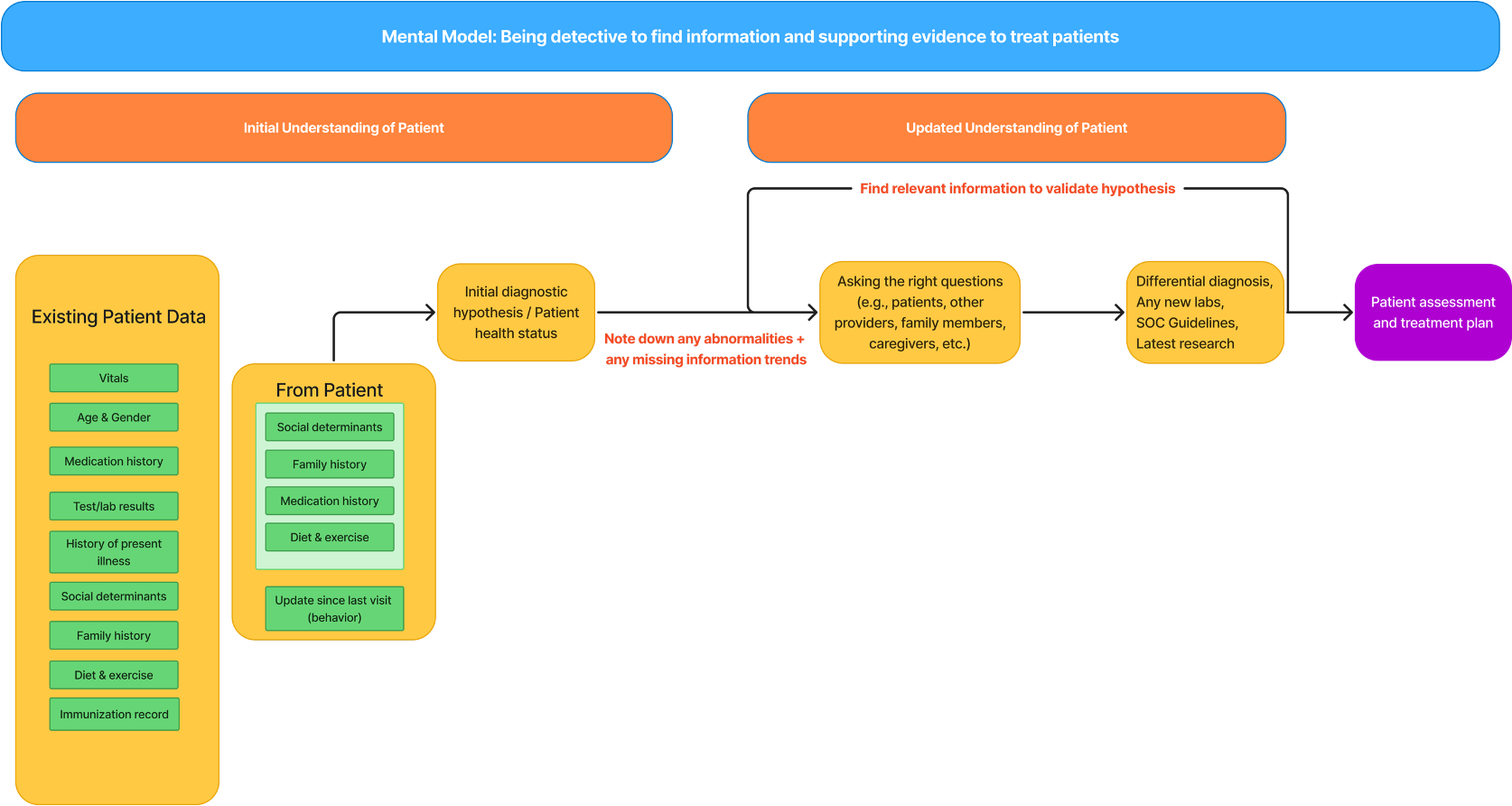}
    \caption{This figure presents physicians' current workflow to collect and analyze patient information throughout patient visits.}
    \label{fig:process}
\end{figure}

\begin{figure}[h!]
    \centering
    \includegraphics[width=400pt]{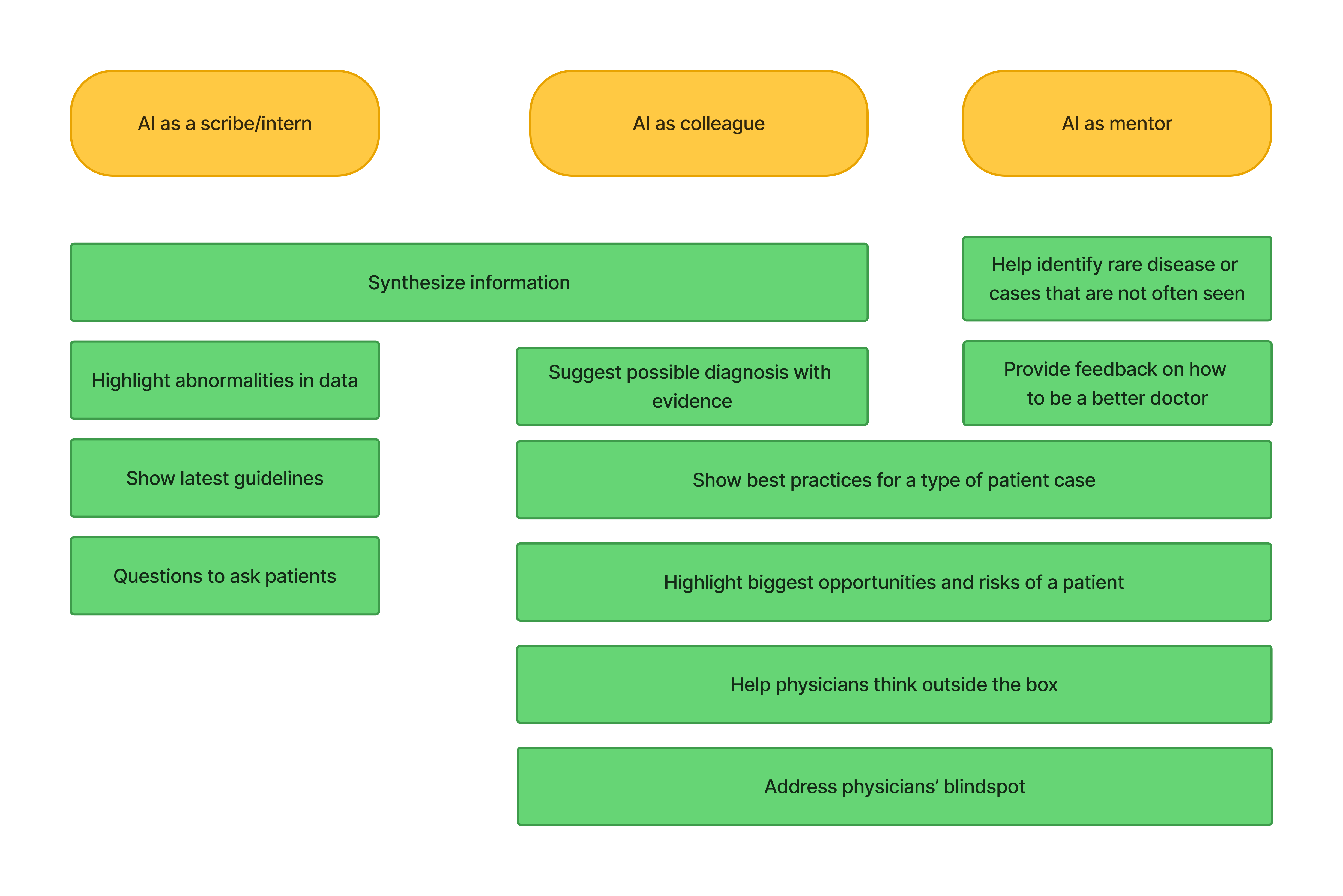}
    \caption{This figure illustrates the informational needs that physicians identified, and the corresponding roles (scribe/intern, colleague, mentor) of the AI should take to address the needs.}
    \label{fig:mentalmodel}
\end{figure}

Physicians often take an investigative stance toward patient data, similar to detectives, assembling and reconciling diverse information sources to form a diagnostic conclusion and treatment plan. Details of their work process are shown in \autoref{fig:process} (Appendix). Physicians in our study expressed informational needs such as: 1) rapidly surfacing critical symptom‑related information from diverse data sources for patient chief complaint(s), 2) conducting a standardized review of electronic medical record systems to ensure comprehensive data collection, and 3) eliciting the appropriate questions/responses to address information gaps. They also described three roles they believe generative AI could serve to help address these needs, including AI as a scribe/intern, as a colleague, and/or as a mentor. The details are illustrated in \autoref{fig:mentalmodel} (Appendix).

\section{Discussion}
In this workshop, we hope to engage in the following conversations.

\subsection{Supporting Physician Information Needs Through Dynamic Exchanges with AI}
Physicians' information needs primarily involve the process of searching, gathering, and synthesizing key patient data and supportive evidence, and how they envision AI dynamically disrupting their workflows. Physicians typically conduct reviews of electronic medical record systems to ensure key information is extracted for patient care. Therefore, they envision AI to search and synthesize all available data related to a patient's chief complaint(s), and for AI to gather information from patients (i.e., workup data and/or responses to questions based on current chief complaints) to update their understanding of a patient's present illness or medical need. They imagine AI presenting information in real-time and supporting multiple input and output modalities (e.g., text, audio, video). They also envisioned AI participating directly in patient encounters, such as surfacing relevant questions as new information emerges and updating electronic medical record systems as needed to improve efficiency. 
\subsection{Generative UI to Align Physicians' Mental Models and Expectations of AI}
As discussed, physicians envisioned multiple roles that AI could play to support their informational needs. These various mental models shaped how they anticipated engaging with an AI system, as well as their expectations for the quality and presentation of its responses. For example, an intern/scribe AI may be expected to provide high-quality, fact-based responses because its task is to synthesize information, and medical decisions will be made based on this synthesis. It is important to create a generative AI user interface system to account for the sheer volume and complexities of medical information situated within contexts such as healthcare stakeholder ecosystems, diseases, socioeconomics, and patients to mitigate errors through design controls. For AI systems envisioned as colleagues or mentors, the interaction design requirements may differ substantially given current technological capabilities and limitations. We therefore aim to further examine the design implications of these roles and explore how generative UI approaches might support the safe and effective integration of AI into clinicians’ information seeking workflows.

\bibliographystyle{ACM-Reference-Format}
\bibliography{sample-base}

\appendix
\section{Appendix}


\end{document}